\documentclass[12pt,preprint]{aastex}

\def\etal{{\it et al.\ }}

\def\cal#1{{\cal #1}}

\def\m@th{\mathsurround=0pt}
\def\n@space{\nulldelimiterspace=0pt \m@th}
\def\biggg#1{{\mbox{$\left#1\vbox to 20.5pt{}\right.\n@space$}}}

\def\beginenum{\begin{enumerate}}
\def\endenum{\end{enumerate}}
\def\bitem{\begin{itemize}}
\def\eitem{\end{itemize}}
\def\bray{\begin{array}}
\def\eray{\end{array}}
\def\begindoc{\begin{document}}
\def\enddoc{\end{document}}
\def\bq{\begin{equation}}
\def\eq{\end{equation}}
\def\bqy{\begin{eqnarray}}
\def\eqy{\end{eqnarray}}
\def\bqyn{\begin{eqnarray*}}
\def\eqyn{\end{eqnarray*}}
\def\bc{\begin{center}}
\def\ec{\end{center}}
\def\bfll{\begin{flushleft}}
\def\efll{\end{flushleft}}
\def\bflr{\begin{flushright}}
\def\eflr{\end{flushright}}

\newcommand{\Avec}{\mbox{\boldmath $A$}}
\newcommand{\Bvec}{\mbox{\boldmath $B$}}
\newcommand{\Evec}{\mbox{\boldmath $E$}}
\newcommand{\Fvec}{\mbox{\boldmath $F$}}
\newcommand{\Gvec}{\mbox{\boldmath $G$}}
\newcommand{\Rvec}{\mbox{\boldmath $R$}}
\newcommand{\Uvec}{\mbox{\boldmath $U$}}
\newcommand{\Vvec}{\mbox{\boldmath $V$}}
\newcommand{\evec}{\mbox{\boldmath $e$}}
\newcommand{\jvec}{\mbox{\boldmath $j$}}
\newcommand{\kvec}{\mbox{\boldmath $k$}}
\newcommand{\nvec}{\mbox{\boldmath $n$}}
\newcommand{\uvec}{\mbox{\boldmath $u$}}
\newcommand{\vvec}{\mbox{\boldmath $v$}}
\newcommand{\wvec}{\mbox{\boldmath $w$}}
\newcommand{\xvec}{\mbox{\boldmath $x$}}
\newcommand{\omegavec}{\mbox{\boldmath $\omega$}}
\newcommand{\Omegavec}{\mbox{\boldmath $\Omega$}}

\begin{document}
\title{Acceleration of Plasma Flows in the Solar Atmosphere Due to
Magnetofluid Coupling - Simulation and Analysis  }
\author{Swadesh M. Mahajan\altaffilmark{1}}
\affil{Institute for Fusion Studies, The University of Texas at  Austin,
Austin, Texas 78712}
\author{Nana L. Shatashvili\altaffilmark{\ 2}
\affil{Plasma Physics Department, Tbilisi State University,
Tbilisi 380028, Georgia\\
International Center for Theoretical Physics, Trieste, Italy}
Solomon V. Mikeladze and Ketevan I. Sigua} \affil{Andronikashvili
Institute of Physics, Georgian Academy of Sciences, Georgia}
\altaffiltext{1}{\small Electronic mail: \
mahajan@mail.utexas.edu} \altaffiltext{2}{\small Electronic mail:
\ shatash@ictp.trieste.it \hskip 0.3cm nanas@iberiapac.ge}

\clearpage

\begin{abstract}
{Within the framework of a two--fluid description  possible
pathways for the generation of fast flows (dynamical as well as
steady) in the lower solar atmosphere is established. It is shown
that a primary plasma flow (locally sub--Alfv\'enic) is
accelerated when interacting with emerging/ambient arcade--like
closed field structures. The acceleration implies a conversion of
thermal and field energies  to kinetic energy of the flow. The
time--scale for creating   reasonably fast flows \ ($\gtrsim
100$\,km/s) \ is dictated by  the initial ion skin depth while the
amplification of the flow depends on local \ $\beta $. It is
shown, for the first time, that distances over which the flows
become "fast"  are   \ $\sim 0.01\,R_s$ \ from the interaction
surface; later the fast flow localizes (with dimensions \
$\lesssim 0.05\,R_S$) \ in the upper central region of the
original arcade. For fixed initial temperature the final speed \
($\gtrsim 500\,km/s$) \ of the accelerated flow, and the
modification of the field structure are independent of the
time-duration (life--time) of the initial flow. In the presence of
dissipation, these flows are likely to play a fundamental role in
the heating of the finely structured Solar atmosphere. }
\end{abstract}


\clearpage


\section{Introduction}
\label{sec:intro}
\bigskip

In astrophysics (particularly  in the physics of the solar
atmosphere), plasma "flow" could be assigned at least two
connotations: 1) The flow is a primary object whose dynamics bears
critically on the phenomena under investigation. The problems of
the formation and the original heating of the coronal structure,
the creation of channels for particle escape, for instance, fall
in this category, 2) The flow is a secondary feature of the
system, possibly created as a by--product and/or used to drive or
suppress an instability. Since the generation of flows which will
eventually create the coronal structures \cite{MMNS1,MMNS2} is the
theme of this effort, the flows  here are fundamental.

By exploiting a simple two--fluid model in the solar context,
several recent studies \cite{MY-1,YOM} have  revealed the breadth
of phenomena made possible by the combined action of the
flow--velocity and the magnetic fields. The flow-based approach
will prove, perhaps, crucial in the study of solar corona,
observationally found to be a highly dynamic arena replete with
multiple--scale spatiotemporal structures (Aschwanden \etal
2001a); the approach gains immense credibility with the discovery
that strong flows are found everywhere -- in the subcoronal
(chromosphere) as well as in the coronal regions  (see e.g.
\cite{schrijver,golub,A1,A2,ami1,ami2,motions,df} and references
therein). Recent phenomenology strongly emphasizes  that the solar
atmosphere is an extremely inhomogeneous (in all parameters) area
in which small-- and large--scale closed magnetic field structures
with different temperatures  co--exist in nearby regions. For
example, two-temperature coronal models constructed from {\it
SOHO/EIT} observations indicate complicated magnetic topology and
fine--scale structuring of corona (including Coronal Holes)
\cite{zhang,chertok}. It is also clear that the mechanisms for
energy transport and channeling of particles in Solar atmosphere
are deeply connected with the challenging and exciting problems of
the solar coronal heating and of the origin of the solar wind (SW)
\cite{woo2}.

If flows are to play an important and essential role in
determining the dynamics and structure of the solar corona, we
must immediately face the problem of finding sources and
mechanisms for the creation of these flows. Catastrophic models of
flow production in which the magnetic energy is suddenly converted
into bulk kinetic energy (and thermal energy) are rather
well--known; various forms of magnetic reconnection (flares, micro
and nano--flares) schemes permeate the literature (E.g.~(Wilhelm
2001; Christopoilou, Georgakilas and Koutchmy 2001) for
chromosphere up--flow generations). A few other  mechanisms of
this genre also exist: Uchida \etal (2001) proposed that the major
part of the supply of energy and mass to the active regions of the
corona may come from a dynamical leakage of magnetic twists
produced in the sub-photospheric convection layer; Ohsaki {\it et
al.} (2001, 2002) have shown how a slowly evolving closed
structure (modelled as a double--Beltrami two--fluid equilibrium)
may experience, under appropriate conditions,  a sudden loss of
equilibrium with the initial magnetic energy appearing as the mass
flow energy. Another mechanism, based on loop interactions and
fragmentations and explaining the formation of loop threads, was
given in Sakai and Furusawa (2002); the suggestion based on
cascade of shock wave interactions was made in \cite{ryta}. A more
quasi-static mechanism for flow generation in sub--coronal regions
taking into account the density in--homogeneity of the structures
was given in \cite{mnsy}.

Before we embark on delineating the flow-generation mechanisms, we
present additional evidence/speculation on their existence as well as
their possible role in the processes taking place in the solar
vicinity:

1) Goodman (2001) has shown that the mechanism which transports
mechanical energy from the convection zone to the chromosphere (to
sustain its heating rate) could also supply the energy needed to
heat the corona, and accelerate the SW. The coronal heating
problem, hence, is shifted to the problem of the dynamic
energization of the chromosphere. In the latter process the role
of flows is found to be critical as warranted by the following
observations made in soft X--rays and extreme ultraviolet (EUV)
wavelengths, and recent findings from the {\it Transition Region
and Coronal Explorer (TRACE)}: the over--density of coronal loops,
the chromospheric up--flows of heated plasma, and the localization
of the heating function in the lower corona (Schrijver, \etal
1999; Aschwanden \etal 2001a; Aschwanden 2001b).

2) The connection/coupling of transient events like up--flows and
different type jet--like structures to the photosphere dynamics
was reported in numerous studies (see e.g. \cite{ryta} and
references in). In \cite{flare} it was demonstrated that the
eruptions of coronal mass ejection is triggered from the low solar
atmosphere (photosphere/chromosphere) as seen in TRACE $1600 A^o$
images and with SOHO Michelson Doppler Imager. The data of this
latest research favor the idea that a catastrophic loss of MHD
equilibrium can be the primary driving mechanism for the rapid
ejection that has 3 important stages - a relatively stable
equilibrium, a loss (fast, impulsive) stage, and the final rapid
eruption (associated with  substantive changes in the photospheric
magnetic flux and white--light morphology). The results of
\cite{sa3} suggest the coupling between magnetic fields and
convective processes that pervades the solar photosphere. The
correlation between photospheric shear flows and flares is also
reported in \cite{rimmele}; several current models suggest that
the former can be responsible for the energy build up of  the
flares.

3) In \cite{sep}, the authors report on the low coronal signatures
of major solar energetic particle (SEP) events focusing on
flare--associated motions (observed in  soft X--rays). It was
underlined that these motions may provide an important link
between small--scale energy release and large--scale explosive
events; the existence of a continuum of acceleration timescales
was pointed out. In \cite{injection} the detailed investigation of
the dynamical behavior of emerging magnetic flux using
three--dimensional MHD numerical simulation was carried out and it
was shown that the emergence generates not only vertical but also
horizontal flows in the photosphere, both of which contribute to
the injection of the magnetic energy and helicity. The
contributions of vertical flows are dominant at the early phase of
flux emergence, while horizontal flows become a dominant
contributor later. In \cite{heatingB} it was shown that solar
corona is mainly heated by the magnetic activity in the edges of
the network flux clumps that are observed to be riddled with the
fine-scale explosive events. They present that: (1) at the edges
of the network flow clumps there are many transient sheared-core
bipoles of the size and lifetime of granules and having transverse
field strengths greater than $\sim 100\,G$, (2) $\sim 30$ of these
bipoles are present per supergranule, and (3) most spicules are
produced by explosions of these bipoles.

4) Recent observations also suggest that the energy for coronal
heating is very likely a by--product of the outflow of heat from
Sun's interior through the convection zone -- the convection zone
acts as a heat engine, converting some of the thermal energy into
mechanical and magnetic energy, some of which enters the corona
and dissipates into heat. There are only two obvious energy
sources that could power significant flow generation in the
chromosphere: the magnetic field (both large scale and
short--scale including turbulence), and the thermal pressure of
the plasma. We have already mentioned a few examples of the
magnetically driven transient, but sudden flow--generation. A more
quiescent pathway was studied in \cite{mnsy} showing the
possibility of magneto--fluid rearrangement of a relatively
constant kinetic energy (going from an initial
high-density--low-velocity state to a low-density--high-velocity
stage). The mechanisms based on the wave--energy transformation
and instabilities can be operative at later stages of the flow
evolution; these mechanisms could have additional importance for
acceleration \cite{prm}.

The main message then, is that to solve the coronal heating
problem, the inclusion of processes  taking place in the
chromosphere and the transition region may be essential. In
particular, one  must take into account the different time--scale
dynamical stages of the evolution of the primary flow as it passes
through specific regions of solar atmosphere areas nested by
varying scale ambient magnetic field structures. The dynamics of
the flow must be understood.

In present paper we will show the possibility of flow
acceleration/generation in the Solar atmosphere based on the
dynamical two--fluid model suggested in \cite{MMNS2}. We will show
that there exists an extremely fast stage (right at the lower
chromosphere heights) giving rise to  significant flow
acceleration/generation; it is followed by a quasi--static stage
in which  the created fast flows are further accelerated via the
magnetofluid coupling (depending on the region of the atmosphere
the density could be constant or spatially varying). The detailed
nature of accelerated flows will depend on the initial and
boundary conditions.


\section{Model}
\label{sec:Model}

The physical model for flow generation/acceleration is a
simplified two--fluid model. The plasma is quasi--neutral  ---
electron and proton number densities are nearly equal: $n_e\simeq
n_i = n$ ($\nabla \cdot {\bf j} = 0$), but  the electron and the
proton flow velocities are allowed to be different. Neglecting
electron inertia, these are $V_i$ and $V_e=(V-j/en)$,
respectively. We assign equal temperatures to the electron and the
protons so that the kinetic pressure $p$ is given by: $p = p_i +
p_e \simeq 2\,nT , \ T = T_i\simeq T_e $. The analysis can be
readily extended later to the more realistic case of different
temperatures for different species \cite{swT}. We understand that
when solving the solar wind problem one should use the
multi--fluid, multi--dimensional descriptions (see e.g.
\cite{marsch1,marsch2,hollweg} and references therein) but we
however, believe, that essential features of the primary
flow--based physics of its acceleration can be captured with our
basic model. Very near the photospheric surface, the influence of
neutrals and ionization (and processes of flux emergence etc.)
would not permit just two--fluid approach. A little farther
distance ($\Delta r \ge 500\,$km) from the surface, however, we
expect that there exist fully ionized and magnetized plasma
structures such that the dynamical two--fluid model will capture
the essential physics of flow generation.

The dimensionless two--fluid equations describing the flow--field
interaction processes can be read from (Mahajan {\it et al.} 1999,
2001):
\begin{equation}
\frac{\partial }{\partial t}{\bf V}+({\bf V}\cdot \nabla ){\bf V}=\frac{1}{n}%
\nabla \times {\bf b}\times {\bf b}-\beta_0\,\frac{1}{n}\nabla
(nT)+\nabla
\left( \frac{r_{A0}}{r}\right) +\nu _{i}(n,T)\left( \nabla ^{2}{\bf V}+\frac{1%
}{3}\nabla (\nabla \cdot {\bf V})\right) ,  \label{eq:motion}
\end{equation}
\begin{equation}
\frac{\partial }{\partial t}{\bf b}-\nabla \times \left( {\bf V}-\frac{%
\alpha_0 }{n}\,\nabla \times {\bf b}\right) \times {\bf
b}=\alpha_0\,\beta_0 \ \nabla \left( \frac{1}{n}\right) \times
\nabla (nT), \label{eq:field1}
\end{equation}
\begin{equation}
\nabla \cdot {\bf b}=0,  \label{eq:field2}
\end{equation}
\begin{equation}
\frac{\partial }{\partial t}n+\nabla \cdot n{\bf V}=0,
\label{eq:cont-1}
\end{equation}
\begin{eqnarray}
\frac{3}{2}n\frac{d}{dt}(2T)+\nabla ({\bf q}_{i}+{\bf q}_{e}) &=&-2nT\nabla
\cdot {\bf V}+2\beta_0^{-1}\,\nu _{i}(n,T)\,n\left[ \frac{1}{2}\left( \frac{%
\partial V_{k}}{\partial x_{l}}+\frac{\partial V_{l}}{\partial x_{k}}\right)
^{2}-\frac{2}{3}\,(\nabla \cdot {\bf V})^{2}\right]  \nonumber \\
&&+\frac{5}{2}\alpha_0\,(\nabla \times {\bf b})\cdot \nabla
T-\frac{\alpha_0}{n}\,(\nabla \times {\bf b})\nabla
(nT)+E_{H}-E_{R}. \label{eq:heat}
\end{eqnarray}
where the notation is standard with the following normalizations:
the density $n$ to $n_{0}$ at some appropriate distance from the
solar surface, the magnetic field to the ambient field strength at
the same distance, and velocities to the Alfv\'{e}n velocity
$V_{A0}$. The parameters $r_{A0}=GM_{\odot }/V_{A0}^{2}R_{\odot
}=2\beta _{0}/r_{c0},\ \alpha _{0}=\lambda _{i0}/R_{\odot },\
\beta _{0}=c_{s0}^{2}/V_{A0}^{2}$ are defined with $n_{0},\
T_{0},\ B_{0}$. Here $c_{s0}$ is a sound speed, $R_{\odot }$ is
the solar radius, $r_{c0}=GM_{\odot }/2c_{s0}^{2}R_{\odot }$,
$\lambda _{i0}=c/\omega _{i0}$ is the collisionless skin depth,
$\nu _{i}(n,T)$ is ion kinematic viscosity and $q_{e}$ and $q_{i}$
are electron and ion dimensionless heat flux densities, $E_{H}$ is
the local mechanical heating function and $E_{R}$ is the total
radiative loss. We note that the full viscosity tensor relevant to
a magnetized plasma is rather cumbersome, and we do not display it
here. Just to have a feel for the importance of spatial variation
in viscous dissipation, we display its relatively simple symmetric
form. It is to be clearly understood that this version is meant
only for theoretical elucidation and not for detailed simulation.
We also note that in general, Hall current contributions are
expected to become significant when the dimensionless Hall
coefficient $\alpha_0 $ satisfies the condition: $\alpha_0 > \eta
$, \ where $\eta $ is the inverse Lundquist number for the plasma.
For a typical coronal plasma as well as for low chromosphere and
transition region (TR) this condition is easily satisfied
($\alpha_0 $ is in the range $10^{-10} - 10^{-7}$ for densities
within $(10^{14} - 10^8) \,cm^{-3}$ and $\eta = c^2/(4\pi
V_{A0}R_{\odot}\sigma ) \sim 10^{-14}$, where $\sigma $ is the
plasma conductivity).

\bigskip

To establish the relevant parameter regime, we resort to recent
observational data (e.g. \cite{goodman,A1,sa} and references
therein). At $\sim (500-5000)\,km$, the observations yield the
average plasma density and temperature to be respectively
$n\sim(10^{14}-10^{11})$cm$^{-3}$ , and \ $T\sim (1-6)\,$eV. For
simplicity, we have assumed $T_e=T_i=T$. The information about the
magnetic field is hard to extract due to the low sensitivity and
lack of high spatial resolution of the measurements coupled with
the inhomogeneity and co--existence of small-- and large--scale
structures with different temperatures, (observational evidence of
small scale mixtures of weak and strong fields \cite{sal}) in
nearby regions. The observation of pixel--to--pixel variations in
the magnetic field indicates that small--scale (sub--pixel)
distribution of fields changes considerably at larger spatial
scales \cite{sa2,sa4}. At these distances we have different values
for the network and for the internetwork fields: (i) The {\it
network} plasmas have typically {\it short-scale} fields in the
range $B_0\sim (700-1500)\,$G, have more or less $n\sim const$ .
(ii) The {\it internetwork fields} are generally weaker (with some
exceptions \cite{sa}) --- $B_o\le 500\,$G, and are embedded in
{\it larger--scale plasma structures} with $n\neq const$ . For
different classes of magnetic field structures different scenarios
may be operative.


\section{Acceleration of particle flows -- analysis of the stages}

In our investigation we shall assume that the processes that
generate the primary flows and the primary solar magnetic fields
are  independent. The  plasma flows begin to interact with
the ambient field at time t=0. The choice of initial conditions
for our numerical work is  guided by the observational evidence
presented in the introduction. Our approach is consistent
with that of Woo, Habbal and Feldman (2004) who have argued that
the flow of the solar wind is influenced by the closed field
structures stressing the self--consistent process of acceleration
and trapping/heating of plasma particles in the finely structured
atmosphere. We will dwell, in this paper, on the representative
problem of the trapping and acceleration of the  primary flow
impinging on a single closed-line structure.  The simulation was
performed for a variety of initial and boundary conditions and
essential aspects of the typical  results will be presented below.


\subsection{Dynamical generation of fast flows}

The general set of Eqs. (\ref{eq:motion}--\ref{eq:heat}) was
solved numerically in Cartesian Geometry for 2.5 Dimensions
($\partial_y=0$). Note that the 2.5D Cartesian nature of our code
does not allow us to explore large distances from the surface due
to interference with the boundaries. Fortunately that does not
translate into a serious shortcoming because much of the action is
found to be limited to regions very close to the surface; the
simulation results, therefore, are quite trustworthy in the
revelation of the basic processes of interest. In carrying out the
simulations an important assumption was made: the diffusion time
of magnetic field is longer than the duration of the interaction
process.

A short summary of our numerical methods is in order. We use the
2.5D version of Lax--Wendroff finite difference numerical scheme
along with applying the Flux--Corrected--Transport procedure
\cite{rm,zalesak}. The predictor-corrector type of approximation
was used. Equation (\ref{eq:field2}) was replaced by its
equivalent for the y--component of the vector potential to ensure
the divergence--free property of the magnetic field. The equation
of heat conduction was treated separately by the alternate
direction implicit method with iterations. Transport coefficients
for heat conduction and viscosity are taken from Braginski, 1965.
In the code, the Bremsstrahlung radiation accounts for $E_R$
\cite{MMNS2}, using a somewhat modified formula assuming it to be
2 times greater, $E_R=2\cdot E_{Br}=2\cdot 1.69 \cdot
10^{-25}\cdot n^2\cdot T^{1/2}\cdot Z^3\,erg/cm^3\,s, \ (Z=1)$.
Since we were exploring a particular heating mechanism suggested
there, no external heating source $E_H$ was needed. A numerical
mesh of $280\times 220$ points was used for computation. The
corresponding scheme is characterized by  second order accuracy
with respect to the chosen grid.

\bigskip

The latest observations support the idea that the coronal material
is injected discontinuously (in pulses or bunches, for example)
from lower altitudes into the regions of interest (e.g., spicules,
jet--like structures). A realistic simulation, then, requires a
study of the interaction of both temporally and spatially
nonuniform initial flows with arcade--like magnetic field
structures. These "close to the actual" cases represent more
vividly the dynamics of the flow acceleration process. Below we
study the dynamics of spatially-temporally  localized flow
(initially a Gaussian, Fig.-s 1,3) entering the region nested with
varying scale arcade--like closed field line structures. For
better visualization of the results we take the symmetric case.
The flow is assumed to be  initially weak \ (${\bf |V|}_{0max}\ll
C_{s0}$). The initial ambient magnetic field was modelled as a
single 2D arcade with circular field lines in the $x$--$z$~plane
(Fig.2 for the vector potential/flux function). The arcade field
attains its maximum value \ $B_{max}(x_o, z=0)\equiv B_{0z}$ \  at
\ $x_0$ \  at its center, and is a decreasing function of the
height $z$ (radial direction). This field was assumed to be
initially uniform in time. When doing so, we choose the parameters
to satisfy the observational constraint that, over a period of
some tens of minutes, the location of the trapping/acceleration
must have a relatively smooth evolution. The final shape and
location of the structure of the associated ${\bf B}({\bf r},t)$,
for example will be naturally defined by its material source, by
the process dynamics, and by the initial field ${\bf B_0}({\bf
r},t)$. We use the following representation for the magnetic
field: \ ${\bf B} = {\bf \nabla} A_y + B_z\,{\bf \hat z}$ \  and
for given geometry \ ${\bf A}(0; A_y; 0); \ {\bf b}={\bf
B}/B_{0z}; \ b_x(t,x,z\neq 0)\neq 0$. From numerous runs on the
flow--field evolution, we have chosen to display pictorially the
results corresponding to the following initial and boundary
conditions: \ $B_{0z}=100\,G$ and flow parameters: \ $V_{max}(x_o,
z=0)=V_{0z}=2.18\cdot 10^5\,cm/s; \ n_{0max}=10^{12}\,cm^{-3}; \
T(x,z=0)=const=T_0 =10\,eV$. \ The background plasma density \
$n_{bg}=0.2n_{0max}$. \ In simulations \ $n(x,z,t=0)=n/n_{0max}$ \
is an exponentially decreasing function of $z$. Experienced gained
after numerous runs, revealing that the processes under study are
localized within a small area of interaction, we settled on the
following boundary condition, \ ${\partial}_x{\mathcal K}(x=\pm
\infty,z,t)=0$ \ which was used with sufficiently high accuracy
for all parameters \ ${\mathcal K}({\bf A},T,{\bf V},{\bf B},n)$ .
Guided by observations we assume that the initial  velocity field
has a pulse--like distribution (Fig.3) with a time duration
(life--time) $t_0$ $\gtrsim 50\,s$.

To illustrate the acceleration of initial flows (extremely weak),
we have modelled several cases with different initial and boundary
conditions. The dynamical picture is strongly dependent on the
relative strengths of the initial flow pressure and the magnetic field
strength.

Our  typical representative  example is the evolution of a
symmetric weak up--flow with its peak located in the central
region of a single closed magnetic field structure (location of
field maximum $B_{0z}=100\,G$) (Figs. 1-3). Figs. (4-8),  in which
we give the $x-z$ contour plots of all the relevant fields ($A_y;
\ |{\bf b}|; \ n; \ |{\bf V}|; \ T$),  contain the essence of the
simulation. We find that the acceleration is significant in the
vicinity of the field--maximum with strong deformation of field
lines and energy re--distribution. In this very region, the
simulations show cooling of the flow with serious density
redistribution: part of the flow is trapped in the maximum field
localization area, accumulated, cooled and accelerated. The
accelerated flow reaches $\gtrsim 100$\,km/s value in less than
$100$\,s (in agreement with recent observations
\cite{ryta,ami1,schrijver} and references therein). The
accelerated flow is decoupled from the mother flow, and is
localized in a distinguishable region with dimensions $\lesssim
0.05\,R_s$ starting at a distance  $\sim 0.01\,R_s$ from the
interaction surface. The time for reaching the quasi--equilibrium
parameters is determined by the initial and boundary conditions
(this conclusion is general for all cases).

In a stationary analysis  to be presented in the next
sub--section, we will attempt to derive the characteristic steady
state parameters (like the distance from the surface)  of the
simulated system.

Extensive simulation experiments show that, when viscosity and
heat flux effects are included, the   flow acceleration
evolution parameters depend strongly on $\alpha_0 $, the
parameter measuring the strength of the Hall term in the two-fluid
equations. A very interesting and far-reaching  result is that the
final parameters of the accelerated flow are practically
independent of the initial flow--characteristics (Fig.-s 4-8);
only the initial fast stage of acceleration up to $\sim 200\,km/s$
is slightly different for different primary flows. Simulation
results for 2 different initial life--times of the flow
($t_0=1000\,s$ -- left panel and $t_0=100\,s$ -- right panel in
Fig.-s 4-8) illustrate this feature.

We also found that  at some critical time,  the solutions split
into two parts; all fields, the magnetic (Fig.-s 4,5), the density
(Fig.6), the velocity (Fig.7) and temperature (Fig.8) exhibit
bifurcation. This process persists for different initial
conditions. In Fig.9 we give time evolution plots of  the maximum
values of all fields ($A_y, \ |{\bf b}|,\ b_p,\ b_z,\ n,\ |{\bf
V}|,\ V_p,\ V_z,\ T$) \ for  a pulse--like flow interacting with a
single arcade--structure for different initial life--times ($t_0$)
of the flow ($t_0=100\,s (black); \ t_0=1000\,s (red); \
t_0=2000\,s (blue); \ t_0\to \infty $ (green)). In Fig.10 the same
maximum values of all fields are plotted versus the initial
life--time ($t_0$) of  the flow for different time--frames
($t=200\,s
(black); \ 500\,s (red); \ 1000\,s (blue); \ 1500\,s (green); \\
2000\,s (light green);\  3000\,s$ (rose)).

These pictures clearly demonstrate  that the accelerated mother
flow bifurcates into 2 separate fast daughter--flows (after an
initial acceleration stage) modifying significantly the original
arcade structure. The characteristic fields undergo rather similar
dynamics for flow pulses with different initial life--times. It
should be emphasized that now the maxima of these parameters are
localized not along the initial B--maximum but on both sides of it
and shifted along height (in the localization areas of each
accelerated daughter flow with newly created B--maxima and
currents). After  the initial acceleration stage, the magnetic
energy maxima remain practically unchanged up to some "blow--up"
time ($\gtrsim 2000\,s$) at which the gradients become too steep
and the  simulation results cease to be meaningful. The same
result holds for the maxima of the transverse and parallel
magnetic field energies (with $\lesssim 10\%$ accuracy). For fixed
$T_0(\beta_0$) the maximum values of each parameter (local in
space) exhibit practically similar dynamics (independent of the
initial flow life--time) reaching similar numbers at
near--critical time. This picture persists for different initial
$T_0(\beta_0$). Testing the conservation of the total energy  of
the system as it evolves in time also shows  that  the simulation
results can be trusted only up to the blow--up time; as one
approaches this time the energy conservation no longer holds. To
study longer time dynamics, the code will need improvement.

We will soon offer a possible explanation of these results through
a simple equilibrium analysis.

\bigskip

We are now in a position to list the most interesting and
distinguishable new results found in a 2.5D simulation of the
two--fluid equations (containing various dissipative and
short--scale effects) solved for different initial and boundary
conditions:
\begin{enumerate}
\item
A primary flow, even with a very slow initial speed \
($V_{0z}\sim 1$\,km/s \,locally sub--Alfv\'enic) is accelerated
when it interacts with an arcade--like closed magnetic field
structure. The effect is strong in the strong field region
(initially the arcade center).  This is a common feature
independent of the arcade--characteristics, and the shape of the
initial flow.
\item
When viscosity and heat--flux are ignored, the
time needed for the flow to acquire reasonable amount of energy is
practically infinite. This is probably due to the fact that
without dissipation, the energy transfer through the short--scales
introduced by the two--fluid effects is not operative unless
special conditions for catastrophic processes pertain.
\item
For
realistic \ $\alpha_0 $ \ (measuring the strength of the Hall
term) the heat flux and viscosity effects cause a re--distribution
of magnetic, flow kinetic and thermal energies in the arcade
region in reasonable times \ $\sim 100$\,s .
\item
During the
redistribution, the arcade field is modified; the thermal-- and
magnetic field-- energies are converted to flow energy locally.
The time--scale for generating a reasonably  fast flow \
($V_{0z}\gtrsim 100$\,km/s) \  is dictated by \ $\alpha_0 $. For a
given initial \ $T_0 (\beta_0)$, the larger the \ $\alpha_0 $ ,
the faster the flow generation. The density is non--uniformly
redistributed within the arcade span.
\item
At some specific
"critical time" \ $\lesssim 1000\,s$ \ the accelerated flow
bifurcates into two separate fast flows. At this moment the arcade
is also split in two, each  with its share of the accumulated
particles. Two fast spicule--like structures, carrying vorticity
and current, are decoupled from the mother flow. Their densities
are similar to the initial density of the mother--flow.
\item
Initially the amplification of the flow depends on the flow \
$\beta_0$ , the ratio of the thermal  and the magnetic field
energy.
\item
The distance from  the  interaction surface where
the bifurcation occurs is \ $\sim 0.01\,R_s$ \,. It is interesting
to mention that this height is lower than the heights of the base
of a typical hot coronal structure (\cite{schrijver,MMNS2} and
references therein) and it seems to be comparable to the latest
observational findings \cite{df,woo2}. Initially the fast flow
localizes in the center of the original arcade, starting from this
distance. After the bifurcation several flows appear with
distinguishable dimensions \ ($\lesssim 0.05\,R_s$) \ practically
on similar heights.
\item
For fixed initial $T_0$, the final speed
of accelerated flow and the picture of the modified field
structure are independent of the initial flow life--time. This
result seems extremely important in connection with the observed
flows in the lower atmosphere. At \ $t\gtrsim 1000$\,s \
velocities reach \ $\sim 500$\,km/s \ or even  greater \
($\lesssim 800$\,km/s) \ values. Such result persists for
different \ $T_0(\beta_0)$ .
\end{enumerate}
We note here that at any quasi--equilibrium stage of the
acceleration process, the nascent intermittent flows will blend
and interact with pre--existing varying scale closed field
structures (recall the fine structure of the solar atmosphere);
the ``new'' flows could be trapped by other structures with
strong/weak magnetic fields participating in creating different
dynamical scenarios: heating of the new structure \cite{MMNS2}
could result, or an escape channel could be created
\cite{channel,woo2}. Instabilities, generation of  waves could
also be triggered.


\subsection{Quasi--equilibrium pathway of flow acceleration due
to magneto--fluid coupling  --- restrictions and analysis}

Both the observational evidence and the results of dynamical
simulation point out that a typical solar structure  passes
through a quasi--equilibrium stage (possibly even a series of
quasi--equilibria) before it reaches the final explosive or
distortion/deformation stage leading to particle escape. Let us
try to understand the physics of these  quasi--equilibrium
structures in terms of  equilibrium two--fluid equations. We
analyze the simplest two--fluid equilibria with \ $T=\rm const
\longrightarrow n^{-1}\nabla p \to \,T\nabla \,ln\,n$ \
(generalization to a homentropic fluid: \ $p=\rm const\cdot
n^{\gamma}$ \ is straightforward and was performed in numerical
work \cite{mnsy}). The dimensionless equations describing the
model equilibrium can be written as:
\begin{equation}
\frac{1}{n}\nabla\times {\bf b\times
b}+\nabla\left(\frac{r_{A0}}{r} -\beta_0\ln\
n-\frac{V^2}{2}\right)+ {\bf V\times (\nabla\times V})=0,
\label{eq:DB-1}
\end{equation}
\begin{equation}
\nabla\times\left[\left({\bf V}-\frac{\alpha_0}{n}\nabla\times{\bf
b} \right)\times {\bf b}\right]=0, \label{eq:DB-2}
\end{equation}
\begin{equation}
\nabla \cdot (n{\bf V})=0, \label{eq:cont}
\end{equation}
\begin{equation}
\nabla\cdot{\bf b}=0, \label{eq:b}
\end{equation}
where ${\bf b}={\bf B}/B_0$ and the following normalizations were
used: $n \to n_0$ -- the density at some appropriate distance from
the solar surface ($\geq 2000\,km$), $B \to B_0 $ -- the ambient
field strength at the same distance, $|V| \to V_{A0}$ and the
dimensionless parameters are defined with $n_0, \ T_0, \ B_0$
taken at the same distance. In the non--dissipative limit, the
system allows the well--known double Beltrami solutions :
\begin{equation}
{\bf b}+\alpha_0 \nabla\times {\bf V}=d\ n\ {\bf V}, \qquad \qquad
{\bf b}=a\ n\ \left[{\bf V}-\frac{\alpha_0}{n}\,\nabla\times {\bf
b}\right], \label{eq:DB-3}
\end{equation}
where $a$ and $d$ are the dimensionless constants related to ideal
invariants: the magnetic $h_1=\int ({\bf A}\cdot {\bf b})\ d^3x$
and the generalized $h_2=\int ({\bf A}+{\bf V})\cdot
\nabla\times({\bf A}+ {\bf V})d^3x$ helicities \cite{MY-1,MMNS2}.
Substituting (\ref{eq:DB-3}) into (\ref{eq:DB-1})--(\ref{eq:cont})
one obtains the Bernoulli Condition
\begin{equation}
\nabla\left(\frac{2\beta_0r_{c0}}{r}-\beta_0\ln\,n-\frac{V^2}{2}\right)=0,\
\label{eq:bernoulli}
\end{equation}
relating the density with the flow kinetic energy, and solar
gravity.

Equations ~(\ref{eq:DB-1}), (\ref{eq:DB-3}),(\ref{eq:bernoulli})
represent a close system. They may be easily manipulated to yield
an alternative form \ ($g(r)=r_{c0}/r$)
\begin{equation}
\frac{\alpha_0^2}{n}\nabla\times\nabla\times{\bf V}+\alpha_0\
\nabla\times \left[\left(\frac{1}{a\, n}-d\right)n\,{\bf
V}\right]+\left(1-\frac{d}{a} \right){\bf V}=0, \label{eq:CC-1}
\end{equation}
\begin{equation}
\alpha_0^2\nabla\times\left(\frac{1}{n}\nabla\times{\bf
b}\right)+\alpha_0\ \nabla \times \left[\left(\frac{1}{a\,
n}-d\right){\bf b}\right]+\left(1-\frac{d}{a} \right){\bf b}=0.
\label{eq:CC-2}
\end{equation}
\begin{equation}
n=\exp\left(-\left[
2g_0-\frac{V^2_0}{2\beta_0}-2g+\frac{V^2}{2\beta_0}
\right]\right). \label{eq:density}
\end{equation}

Equations ~(\ref{eq:CC-1}), and (\ref{eq:density}) can be solved for
the density and the velocity field ${\bf V}$ and then ${\bf b}$ could
be determined from ~(\ref{eq:DB-3}).

In the Solar atmosphere one observes quasi--equilibrium magnetic
structures with both homogeneous (practically anywhere) and
inhomogeneous (especially in the Chromosphere and TR) densities.
By invoking appropriate variational principles, one can show that
the generic double Beltrami class of equilibria are accessible in
all cases of interest: constant density, constant temperature, or
when the plasma obeys an equation of state. Maximum analytical
headway, however, is possible for constant density. In that case
the Bertrami--Bernoulli system consists of a set of linear
equations and has two well--defined scales of variation.
Non--constant density does not lead to a linear chain (see
(\ref{eq:CC-1}), and (\ref{eq:density})), but allows phenomena
peculiar to nonlinear systems. It is the latter class of systems
that we will deal with now.

We will now calculate the amplification conditions for
inhomogeneous density flows in the chromosphere. We restrict to a
one--D variation (along the height $Z$) and choose the constants \
$a\sim d = 100$ \ so that \ $(a-d)/a\,d\sim 10^{-6}$. This choice
insures that two homogeneous Beltrami scales will be vastly
different. Detailed algebraic derivation of the approximate
formulas used below can be found in Appendix 1.

The principal results of Appendix 1 are that if \ $n \gg
(a\,d)^{-1}$ \ (density fall in the region of interest is not more
than 3 orders of magnitude), then

1) the transverse components of magnetic field vary keeping
 $b_x^2+b_y^2=b_{0\perp}^2=const$.

2) The density and the velocity fields are related approximately
by \ $|V|^2=1/d^2n^2$ \ so that the magnetic energy does not
change much, \ $|{\bf b}|^2=const $ \ to leading order.

3) The  Bernoulli condition  transforms to the defining equation
for density:
\begin{equation}
\left(-2\,\beta_0\,n^2+\frac{1}{d^2}\right)\frac{\partial
n}{\partial z}=n^3\,g. \label{eq:flows19-1}
\end{equation}
We notice that for the density to drop with height, it has to be
larger than \ $n_{min} = (2\beta_0)^{1/2} d$. The existence of
$n_{min}$ also implies via \ $V^2=1/d^2n^2$ \ that the maximum
allowed velocity is
\begin{equation}
|V_{max}|=\frac{1}{d\,n_{min}}=(2\beta_0)^{1/2}.
\label{eq:flows20-1}
\end{equation}

As one approaches the singularity at $ n=n_{min}$, the spatial
variation of density (and in particular of the velocity) becomes
very large. In such a region of the steep fall in density and rise
in  velocity, the time--independent dissipationless approach will
not be valid. The Bernoulli equation~(\ref{eq:flows19-1}),
however, clearly reveals the origin of the very fast first stage
of dynamical acceleration found in the  simulations. From
Eq.(\ref{eq:flows19}) we also see that the distance over which the
catastrophe appears is determined by the strength of gravity,
$g(z)$ . Eventual amplification of the flow is determined by the
local value of \ $\beta_0$. These simple consequences of the
Bernoulli equation explain one of the most important findings of
the simulation: for a fixed initial temperature, the final
characteristic parameters of the accelerated flow
(quasi--equilibrium after the fast stage of acceleration) do not
depend on its initial state. For these gross features of the
system, the value of \ $\alpha_0$ \, as long as it is finite, is
also quite irrelevant, it just determines how fast the transverse
components of magnetic and velocity fields oscillate. However when
dissipation is present, \ $\alpha_0$ \, through the mediation of
short-scale physics, plays a crucial role in the
acceleration/heating processes.

In connection with this result it is interesting to mention
that according to latest observations  on the quasi--equilibrium
coronal loops, the so called quasi--periodic intensity oscillations
are found to propagate upwards with  speeds of the order of the
(adiabatic/isothermal) coronal sound speed (\cite{loop-trace} and
the references therein).

\bigskip

For structures with ($n=const$), there are two distinct scenarios
for eruptive events in the current framework : (1) when a "slowly"
evolving structure finds itself in a state of no equilibrium and
(2) when the process of creating a long--lived hot structure is
prematurely aborted; the flow shrinks/distorts the structure that
suddenly shines and/or releases energy or ejects particles. The
latter mechanism requires a detailed time--dependent treatment.
The semi--equilibrium, collisionless magnetofluid treatment
pertains only to the former case \cite{osym1,osym2}. In the
references cited, the conditions for catastrophic transformations
of an  original DB (double Beltrami state) were investigated. It
was shown that when the total energy of the original state exceeds
a critical value, the DB equilibrium suddenly relaxes to a single
Beltrami state corresponding to the larger macroscopic scale; at
the transition, much of the magnetic energy $|{\bf b}|^2$ of the
original state is converted to  heat/flow kinetic energy.


\section{Summary of the results and Conclusions}
\label{sec:conclusions}

We have developed a 2.5 Dimensional dynamical code for two-fluid
equations. The two fluid equations contain the Hall term
($\alpha_0\neq 0 $), the ion vorticity, heatflux and viscosity
effects. We have used the code for a systematic study of particle
acceleration and energy re--distribution phenomena associated with
the interaction of a primary plasma flow with closed field--line
magnetic structures. We also developed simple analytical arguments
to explain and understand essential features of the simulation
results. The simulation and analytical effort have led us to
several far--reaching results for the understanding of the solar
atmosphere.  Even at the cost of some repetition,
we list  the most important ones :\\
(1) A  primary plasma flow (locally sub-Alfv\'enic) is accelerated
when it impinges on  an emerging/ambient arcade--like closed
magnetic field structure. The effect is strong in the strong field
region.  It is found that the final state of the flow is quite
insensitive to the details of initial and boundary conditions; the
latter simply dictate the time--scale at which significant
flow--energy is generated. \\
(2) It is shown that there is a redistribution of magnetic, flow-
kinetic and thermal energies in the arcade region so that the
original arcade field is modified, and thermal and field energies
are converted to flow energy. The time--scale of the fast
($\gtrsim 100$\,km/s) flow generation is dictated by $\alpha_0 $,
the measure of the Hall term . \\
(3) It is found that at some specific time the accelerated flow
bifurcates into 2 separate fast flows with an accompanying split
of the arcade each containing its share of the accumulated particles. \\
(4) Initially the amplification of the flow depends on $\beta_0$
as proven by the 1D analysis; it is shown for the first time that
the distance on which  it happens is $\sim 0.01\,R_s$ (independent
of $\alpha_0$) from the interaction surface. Later the fast flow
localizes (with dimensions $\lesssim 0.05\,R_S$) in the upper
center of the original arcade. \\
(5) It is shown that for fixed initial \ $T_0$ \ the final speed
($\gtrsim 500\,km/s$) of the accelerated flow, and the  shape of
the modified field structure are independent of the initial flow
life--time. Many of these parameters  can be approximately
calculated by analysis.\\

\bigskip

We have shown  possible pathways  for both dynamical and steady
generation of fast flows. The cold flows originating, for example,
in the lower chromosphere acquire energy as they meet and interact
with emerging/ ambient magnetic fluxes; the trapping of an ionized
$\gtrsim 3\,$eV plasma by network/ inter--network structures takes
place at the same time. In the presence of dissipation, these
flows are likely to play a fundamental role in the heating of the
finely structured solar atmosphere.  The explicit purpose of this
paper, however, was to demonstrate the generation of flows in the
lower atmosphere feeding on the ambient magnetic energy. The
flows, in turn, provide  a steady and assured source of matter and
energy for the formation and primary heating of the corona as well
as for the  creation of the solar wind. The agreement of our
preliminary results   with the observation data lends credence and
promise to attempts, based on the exploitation of sub--coronal
flows, to tackle unresolved problems like the coronal heating and
origin of the solar wind. We believe that the chromospheric mass
outflows, spicules, explosive events in chromosphere, micro-- and
nano--flares, large coronal flares, erupting prominences and CMEs
may happen separately but can also be parts of a more global
dynamic process of coronal specific regional formation.

\section*{ACKNOWLEDGEMENTS}

Authors are extremely grateful to Prof. E. Parker for his valuable
discussions and encouragement. Authors thank Dr. E. Marsch and Dr.
K.I. Nikol'skaya for interesting and useful discussions. Authors
express their deepest thanks to Dr. R. Miklaszewski for his
leading suggestions when constructing the improved code. All the
authors thank Abdus Salam International Centre for Theoretical
Physics, Trieste, Italy. The work of SMM was supported by USDOE
Contract No.~DE--FG03--96ER--54366. The work of NLS, SVM and KIS
was supported by ISTC Project G-633.

\clearpage


\appendix{\bf Appendix 1 --- Equilibrium analysis of particle acceleration
for non--uniform density case due to Magneto--fluid coupling}

Let's rewrite DB equations (\ref{eq:DB-3}) in following way:
\begin{equation}
\alpha_0\nabla \times \mbox{\boldmath $b$} = - \frac{1}{a}
\,\mbox{\boldmath $b$} + n\,\mbox{\boldmath $V$}, \qquad \qquad
\alpha_0\nabla \times \mbox{\boldmath $V$} = - \mbox{\boldmath
$b$}  + d \, n\,\mbox{\boldmath $V$}, \label{eq:flows2}
\end{equation}
Let's  define a vector:
\begin{equation}
\mbox{\boldmath $Q$} =  \left(
\begin{array}{c}
\mbox{\boldmath $b$} \\
\mbox{\boldmath $V$}
\end{array}
\right),  \label{eq:flows3}
\end{equation}
then (\ref{eq:flows2}) may be written as:
\begin{equation}
\alpha_0 \nabla \times \mbox{\boldmath $Q $} = M \ \mbox{\boldmath
$Q$}, \label{eq:flows4}
\end{equation}
where
\begin{equation}
M =  \left(
\begin{array}{c}
- a^{-1}\,, \quad n \\
 -1\,, \quad d\,n
\end{array}
\right).  \label{eq:flows5}
\end{equation}
$M$ can be diagonalized by a similarity transformation:
\begin{equation}
S\ M\ S^{-1} = =  \left(
\begin{array}{c}
\lambda_+\,, \quad 0 \\
0\,, \quad \lambda_-
\end{array}
\right),  \label{eq:flows6}
\end{equation}
where \ $[\lambda^2-(d\,n -a^{-1})\lambda +n\,(1-d\,a^{-1})=0]$
\quad $\lambda_{\pm}= \frac{1}{2}[(d\,n-a^{-1})\pm\sqrt{(
d\,n+a^{-1})^2-4\,n}\ ] $ are standard roots. $S$ is found to be
($n$ is a slowly varying parameter, see the Bernoulli condition --
$V^2$ and $g$ are slowly varying):
\begin{equation}
S =  \left(
\begin{array}{c}
1\,, \quad -(\lambda_+ + a^{-1}) \\
1\,, \quad -(\lambda_- + a^{-1})
\end{array}
\right).  \label{eq:flows8}
\end{equation}
Then, if density fall is at a much slower rate than the slow scale
of the Beltrami system (\ $\lambda_-/\alpha_0$, \ ), rewriting
(\ref{eq:flows4}) as:
\begin{equation}
\alpha_0 \nabla \times S\,\mbox{\boldmath $Q $} =
(S\,M\,S^{-1})\,S\, \,\mbox{\boldmath $Q $}= \left(
\begin{array}{c}
\lambda_+ \,, \quad 0 \\
0\,, \quad \lambda_-
\end{array}
\right) \ S\, \mbox{\boldmath $Q$} \ , \label{eq:flows9}
\end{equation}
one finds:
\begin{equation}
S \, \,\mbox{\boldmath $Q $}  = \left(
\begin{array}{c}
\mbox{\boldmath $Q_+$}  \\
\mbox{\boldmath $Q_-$}
\end{array}
\right)= \left(
\begin{array}{c}
\mbox{\boldmath $b$}-(\lambda_+ + a^{-1})\ \mbox{\boldmath $V$}  \\
\mbox{\boldmath $b$}-(\lambda_- + a^{-1})\ \mbox{\boldmath $V$}
\end{array}
\right) \label{eq:flows10}
\end{equation}
each obeying its own independent (fully de--coupled) equation:
\begin{equation}
\nabla \times \mbox{\boldmath $Q_{\pm}$} =
\frac{\lambda_{\pm}}{\alpha_0}\ \mbox{\boldmath $Q_{\pm}$} .
\label{eq:flows11}
\end{equation}
Let's find the amplification conditions for flows (say in the
chromosphere, where $a\sim d = 100$ so that $(a-d)/a\,d\sim
10^{-6}$). Assuming (this is found to be a restriction) \ $n \gg
(a\,d)^{-1}$ \ --- density fall is not more than 3 orders of
magnitude, then
\begin{equation}
\lambda_+ \sim d\,n \ , \qquad \qquad \lambda_-\sim
\frac{a-d}{a\,d} \ . \label{eq:flows13}
\end{equation}
Notice, that for realistic solar atmosphere parameters
(chromosphere, TR, corona) \ $\alpha_0 \sim 10^{-9} - 10^{-11}$ \
and the fast Beltrami scale \  $\lambda_+/\alpha_0\sim 10^{11} -
10^{13}$ \  is very oscillatory and its amplitude must go to zero.
This gives a relation between the velocity and the magnetic field;

\begin{equation}
\mbox{\boldmath $Q_+$}=\mbox{\boldmath $b$} - (d\,n -
a^{-1})\,\mbox{\boldmath $V$}\simeq \mbox{\boldmath $b$} - d\,n
\,\mbox{\boldmath $V$} = 0, \label{eq:flows14}
\end{equation}
and  the approximate equation for the pertinent solution takes the form
\begin{equation}
\nabla \times \mbox{\boldmath $Q_-$} = \frac{a-d}{a\,d\,\alpha_0}
\ \mbox{\boldmath $Q_-$} \qquad \qquad {\rm with} \qquad \qquad
\mbox{\boldmath $Q_-$} = \mbox{\boldmath $b$} -
\frac{\mbox{\boldmath $V$}}{d} \simeq \mbox{\boldmath $b$}.
\label{eq:flows15}
\end{equation}
Let's consider a 1D problem ($Z$ along height, $b_0=1$ when
normalized). Eq.(\ref{eq:flows15}) leads to:
\begin{equation}
\frac{\partial }{\partial z}\left(b_x^2 + b_y^2\right) = 0 \qquad
\Longrightarrow \qquad b_x^2 + b_y^2 = b_{0\perp}^2.
\label{eq:flows17}
\end{equation}
Then, using eq.(\ref{eq:flows14}), one has: \ $V_x^2 + V_y^2 =
b_{0\perp}^2/d^2\,n^2$\ . From Continuity Equation and DB
condition: \ $V_z=V_{0z}/n\sim b_{0z}/d\,n $\ . Thus,
\begin{equation}
V^2=\frac{1}{d^2\,n^2}. \label{eq:flow18}
\end{equation}
Eq.(\ref{eq:flow18}) converts the  Bernoulli condition ($T_0=const$)
to:
\begin{equation}
\left(-2\,\beta_0\,n^2+\frac{1}{d^2}\right)\frac{\partial
n}{\partial z}=n^3\,g. \label{eq:flows19}
\end{equation}
Notice, that maximum allowed velocity for this mechanism is
(compare with the condition (10) of \cite{mnsy}):
\begin{equation}
|V_{max}|=\frac{1}{d\,n_{min}}=(2\beta_0)^{1/2}.
\label{eq:flows20}
\end{equation}
Analysis gives similar results for  varying temperature
($T=n^{-\mu}, \quad 0 <\mu <1$).

\clearpage

\figcaption[Fig.1.eps]{ Initial  symmetric profiles of the radial
velocity $V_z$, and density $n$. The respective maxima (at $x$=0)
are  $\sim 2\,km/s$ and $10^{12}\,cm^{-3}$ . }

\figcaption[Fig.2.eps]{Contour plots for the $y$-- component of
vector potential $A$ (flux function) in the $x-z$ plane for a
typical ambient arcade--like solar magnetic field (initial
distribution). The field has a maximum
$B_{max}(x_0=0,z_0=0)=100$\,G .}

\figcaption[Fig.3.eps]{The original pulse is limited in time. A
time plot of $V_{z,max}(t,z=0)$  corresponding to the shape
$V_z(t,z=0)=V_{0z}\,sin(\pi t/t_0); \  V_z(t>t_0)=0$. The
parameter $t_0$ (1000\,s for this pulse) can be interpreted as the
"life--time" of the pulse.}

\figcaption[Fig.4.eps]{$x-z$ contour plots at various
time--frames: $t=200\,s; \ 500\,s; \ 1000\,s; \ 1500\,s; \\
2000\,s; \ 3000\,s$ \ for the dynamical evolution of $A_y$ for
flows with two different initial life--times. The spatially and
temporally inhomogeneous (type displayed in Fig.1, Fig.3) primary
flows are accelerated as they make their way through the magnetic
field with an arcade--like structure (Fig.2). The realistic
viscosity and heat--flux effects as well as the Hall term
($\alpha_0=3.3\cdot 10^{-10}$ realistic) are included in the
simulation. Left panel corresponds to the case of initial flow
life--time: $t_0=1000\,s$, right panel --- for $t_0=100\,s$. There
is a critical time ($\lesssim 1000\,s$) when the accelerated flow
bifurcates in 2; the original arcade field is deformed
correspondingly.}

\figcaption[Fig.5.eps]{$x-z$ contour plots for the dynamical
evolution of \ $|{\bf b}|$ \ exactly following  the pattern of
Fig.4. After  the bifurcation (read caption of Fig.4),  strong
magnetic field localization areas, carrying currents, are created
symmetrically about $x=0$.}

\figcaption[Fig.6.eps]{$x-z$ contour plots for dynamical evolution
of density \ $n$ \ exactly following  the pattern of Fig.4.
Post--bifurcation daughter flows are localized in the newly
created magnetic field localization areas. The maximum density of
each daughter flow is of the order of the density of the
mother--flow. Daughter--flows have distinguishable dimensions
$\sim 0.05\,R_s$}

\figcaption[Fig.7.eps]{$x-z$ contour plots for the dynamical
evolution of \ $|{\bf V}|$ \ exactly following  the pattern of
Fig.4. The initial flow, locally sub--Alfv\'enic, is accelerated
reaching significant speeds ($\gtrsim 100$\,km/s) in a very short
time ($\gtrsim 100$\,s). The effect is strong in the strong field
region (center of the arcade). At $t\gtrsim 1000$\,s, the
velocities reach $\sim 500$\,km/s or even greater ($\lesssim
800$\,km/s) values. The distance from surface where it happens is
\ $\gtrsim 0.01\,R_s$ . }

\figcaption[Fig.8.eps]{$x-z$ contour plots for dynamical evolution
of temperature $T$ exactly following  the pattern of Fig.4. In the
regions of localization of the daughter flow  there is a
significant cooling while the nearby regions are heated. }

\figcaption[Fig.9.eps]{Dynamical evolution of the characteristic
fields (their maximum values), $|{\bf b}|; \ b_x; \ b_y; \\ n; \
|{\bf V}|; \ V_x; \ V_y; \ T$, defining the interacting
flow--magnetic field system (their $x-z$ contour plots are shown
in Fig.-s 4--8) for different initial flow life--times
($t_0=100\,s \ ({\rm black}); 1000\,s \ ({\rm red}); 2000\,s \ (
{\rm blue}); \infty \ ({\rm green})$). The code ceases to be
dependable for times at which very steep gradients emerge; the
blow--up time for this simulation is ($\lesssim 3000\,s$) }

\figcaption[Fig.10.eps]{Maximum values of  $|{\bf b}|; \ b_x; \
b_y; \ n; \ |{\bf V}|; \ V_x; \ V_y; \ T$ (their $x-z$  contour
plots are\\ shown in Figs. 4--8) versus  the initial life--time
($t_0$) of the primary outflow for different \\ time--frames
($t=200\,s \ ({\rm black}); 500\,s \ ({\rm red}); 1000\,s \ ( {\rm
blue}); 1500\,s \ ({\rm green}); 2000\,s \ ({\rm light green});\\
3000\,s \ ({\rm rose})$). The code ceases to be dependable for
times at which  very steep gradients emerge; the blow--up time for
this  simulation is ($\lesssim 3000\,s$) }


\clearpage

\begin{thebibliography}{99}

\bibitem[Aschwanden {\it et al.} 2001a]{A1} Aschwanden, M.J., Poland
A.I., and Rabin D.M. 2001a, Ann. Rev. Astron. Astrophys., 39, 175.

\bibitem[Aschwanden 2001b]{A2} Aschwanden, M.J. 2001b, \apj, 560,
1035.

\bibitem[Braginski 1965]{braginski} Braginski, S.I. "Transport processes
in a plasma," in Reviews of Plasma Physics, edited by M. A.
Leontovich. Consultants Bureau, New York, (1965), Vol.1, p.205.

\bibitem[Brynildsen {\it et al.} 2004]{df} Brynildsen, N., Maltby,
P., Kjeldseth-Moe, O. \& Wilhelm, K. 2004, \apj, 612, 1193.

\bibitem[Chertok {\it et al.} 2002]{chertok} Chertok, I.M., Mogilevsky, E.I,
Obridko, V. N., Shilova, N. S. and Hudson, H. S. 2002, \apj, 567,
1225.

\bibitem[Christopoilou \& Georgakilas \& Koutchmy 2001]{koutchmy}
Christopoulou, E.B., Georgakilas, A.A., and Koutchmy, S. 2001,
Solar Phys., 199, 61.

\bibitem[De Moortel \& Parnell \$ Hood 2003]{loop-trace} De Moortel,
I., Parnell, C.E. and Hood, A.W. 2003, Solar Phys. 215, 69.

\bibitem[Falconer {\it et al.} 2003]{heatingB} Falconer, D.A., Moore, R.L.,
Porter, J.G. and Hathaway, D.H. 2003, \apj, 593, 549.

\bibitem[Feldman \& Landi \& Curdt 2003]{motions} Feldman, U.,
Landi, E. and Curdt, W. 2003, \apj, 585, 1087.

\bibitem[Goodman 2001]{goodman} Goodman, M.L. 2001, Space Sci. Rev.,
95, 79.

\bibitem[Goodman 2000]{goodman2} Goodman, M.L. 2000, \apj, 533, 501.

\bibitem[Hollweg 1999]{hollweg} Hollweg, J.V., 1999, J. Geophys. Res., 104, 505.

\bibitem[Liu {\it et al.} 2003]{flare} Liu, Y., Jiang, Y., Ji, H.,
Zhang, H. and Wang, H. 2003, \apj, 593, L137.

\bibitem[Magara \& Longscope 2003]{injection} Magara,T. and Longcope, D.W.
2003, \apj, 586, 630.

\bibitem[Mahajan {\it et al.} 1999]{MMNS1} Mahajan, S.M., Miklaszewski,
R., Nikol'skaya, K.I., and Shatashvili N.L., February 1999,
Preprint IFSR \#857, {\it The University of Texas, Austin}, 67
pages.

\bibitem[Mahajan {\it et al.} 2001]{MMNS2}  Mahajan, S.M., Miklaszewski,
R., Nikol'skaya, K.I., and Shatashvili, N.L. 2001, Phys. Plasmas,
8, 1340.

\bibitem[Mahajan {\it et al.} 2002]{mnsy} Mahajan, S.M.,
Nikol'skaya, K.I., Shatashvili, N.L. \& Yoshida, Y. 2002, \apj,
576, L161.

\bibitem[Mahajan {\it et al.} 2003]{channel} Mahajan, S.M., Miklaszewski, R.,
Nikol'skaya, K.I. and Shatashvili, N.L. 2003, ArXive:
astro-ph/0308012.

\bibitem[Mahajan \& Yoshida 1998]{MY-1} Mahajan, S.M., and Yoshida,
Z. 1998, Phys. Rev. Lett., 81, 4863.

\bibitem[McKenzie \& Sukhorukova \& Axword 1998]{swT} McKenzie, J.F.,
Sukhorukova, G.V. and Axword,W.I., 1998, Astronomy and Astrophys.
330, 1145.

\bibitem[Nikol'skaya \& Valchuk 1998]{nv} Nikol'skaya, K.I., and
Valchuk, T.E. 1998, Geomagnetizm and Aeronomy, 38, No.~2, 14.

\bibitem[Nitta \& Cliver \& Tylka 2003]{sep} Nitta, N.V., Cliver, E.
W. and Tylka, A.J. 2003, \apj, 586, L103.

\bibitem[Ohsaki {\it et al.} 2001]{osym1} Ohsaki, S., Shatashvili, N.L.,
Yoshida, Z., and Mahajan, S.M. 2001, \apj, 559, L61.

\bibitem[Ohsaki {\it et al.} 2002]{osym2} Ohsaki, S., Shatashvili, N.L.,
Yoshida, Z., and Mahajan, S.M. 2002, \apj, 570.

\bibitem[Orlando \& Peres \& Serio 1995a]{flows1} Orlando, S., Peres,
G., and Serio, S. 1995a, Astrophys. and Astron., 294, 861.

\bibitem[Orlando \& Peres \& Serio 1995b]{flows2} Orlando, S., Peres,
G., and Serio, S. 1995b, Astrophys. and Astron., 300, 549.

\bibitem[Poedts \& Rogava \& Mahajan 1998]{prm} Poedts, S., Rogava,
A.D. and Mahajan, S. M. 1998, \apj, 505, 369.

\bibitem[Richtmyer \& Morton 1967]{rm} Richtmyer, R.D. and Morton,K.W.
Difference Methods for Initial--Value Problems. Interscience, New
York, 1967.

\bibitem[Ryutova \& Tarbell 2003]{ryta} Ryutova, M. and Tarbell, T.
2003, Phys. Rev. Lett., 90, 191101.

\bibitem[Sakai \& Furusawa 2002]{sakai1} Sakai, J.I., and Furusawa,
K. 2002, \apj, 564, 1048.

\bibitem[Schrijver {\it et al.} 1999]{schrijver} Schrijver, C.J.,
Title, A.M., Berger, T.E., Fletcher, L., Hurlburt, N.E.,
Nightingale, R.W., Shine, R.A., Tarbell, T.D., Wolfson, J., Golub,
L., Bookbinder, J.A., Deluca, E.E., McMullen, R.A., Warren, H.P.,
Kankelborg, C.C., Handy B.N., and DePontieu, B. 1999, Solar Phys.,
187, 261.

\bibitem[Seaton {\it et al.} 2001]{ami1} Seaton, D.B., Winebarger, A.R.,
DeLuca, E.E., Golub, L., and Reeves, K.K. 2001, \apj, 563, L173.

\bibitem[Socas--Navarro \& Sanchez Almeida 2002]{sa} Socas--Navarro,
H., and Sanchez Almeida, J. 2003, \apj, 593, 581.

\bibitem[Socas--Navarro \& Sanchez Almeida 2003]{sa2} Socas--Navarro,
H., and Sanchez Almeida, J. 2002, \apj, 565, 1323.

\bibitem[Socas--Navarro \& Martinez Pillet \& Lites 2004]{sa3} Socas--Navarro,
H., Martinez Pillet, V. and Lites, B.W. 2004, \apj, 611, 1139.

\bibitem[Socas--Navarro 2004]{sa4} Socas--Navarro, H. 2004, \apj, 613, 610.

\bibitem[Socas--Navarro \& Lites 2004]{sal} Socas--Navarro, H. and
Lites, B.W. 2004, \apj, 616, 587.

\bibitem[Tu \& Marsch 1997]{marsch1} Tu, C.-Y. and Marsch, E., 1997,
Solar Phys., 171, 363.

\bibitem[Tu \& Marsch 2001]{marsch2} Tu, C.-Y. and Marsch, E. 2001, J.
Geophys. Res., 106, 8233.

\bibitem[Uchida {\it et al.} 2001]{uchida} Uchida Y., Miyagoshi, T., Yabiku
T., Cable S., and Hirose S. 2001, Publ. Astron. Soc. Japan, 53,
331.

\bibitem[Wilhelm 2001]{wilhelm} Wilhelm, K. 2001, Astrophys. and
Astronomy, 360, 351.

\bibitem[Winebarger \& DeLuca \& Golub 2001]{golub} Winebarger, A.M.,
DeLuca, E.E., and Golub, L. 2001, \apj, 553, L81.

\bibitem[Winebarger {\it et al.} 2002]{ami2} Winebarger, A.R., Warren,
H., Van Ballagooijen, A., DeLuca E.E., and Golub, L. 2002, \apj,
567, L89.

\bibitem[Woo \& Habbal \& Feldman 2004]{woo2} Woo, r, Habbal, S.R. \& Feldman,
U. 2004, \apj, 612, 1171.

\bibitem[Zalesak 1979]{zalesak} Zalesak, S.T. 1979, J. Comput. Phys. 31, 335.

\bibitem[Zhang {\it et al.} 1999]{zhang} Zhang, J., White, S.M.
and Kundu, M.K. 1999, \apj, 527, 977.

\bibitem[Yang {\it et al.} 2004]{rimmele} Yang, G., Xu, Y., Cao,
W., Wang, H., Denker, C. and Rimmele, T.R. 2004, \apj, 617, L151.

\bibitem[Yoshida \& Ohsaki \& Mahajan 2004]{YOM} Yoshida, Z., Ohsaki, S.
and Mahajan, S.M. 2004, Phys. Plasmas, 11, 3660.

\end{thebibliography}
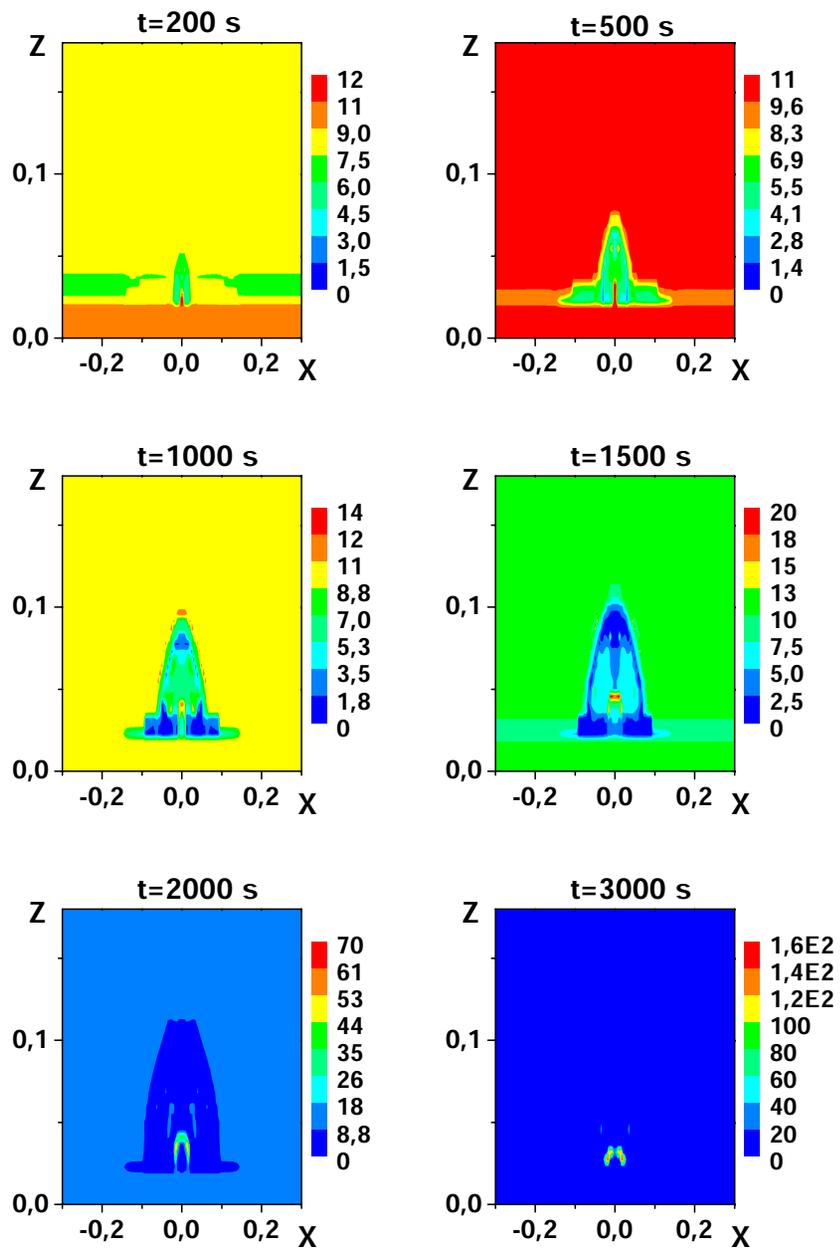
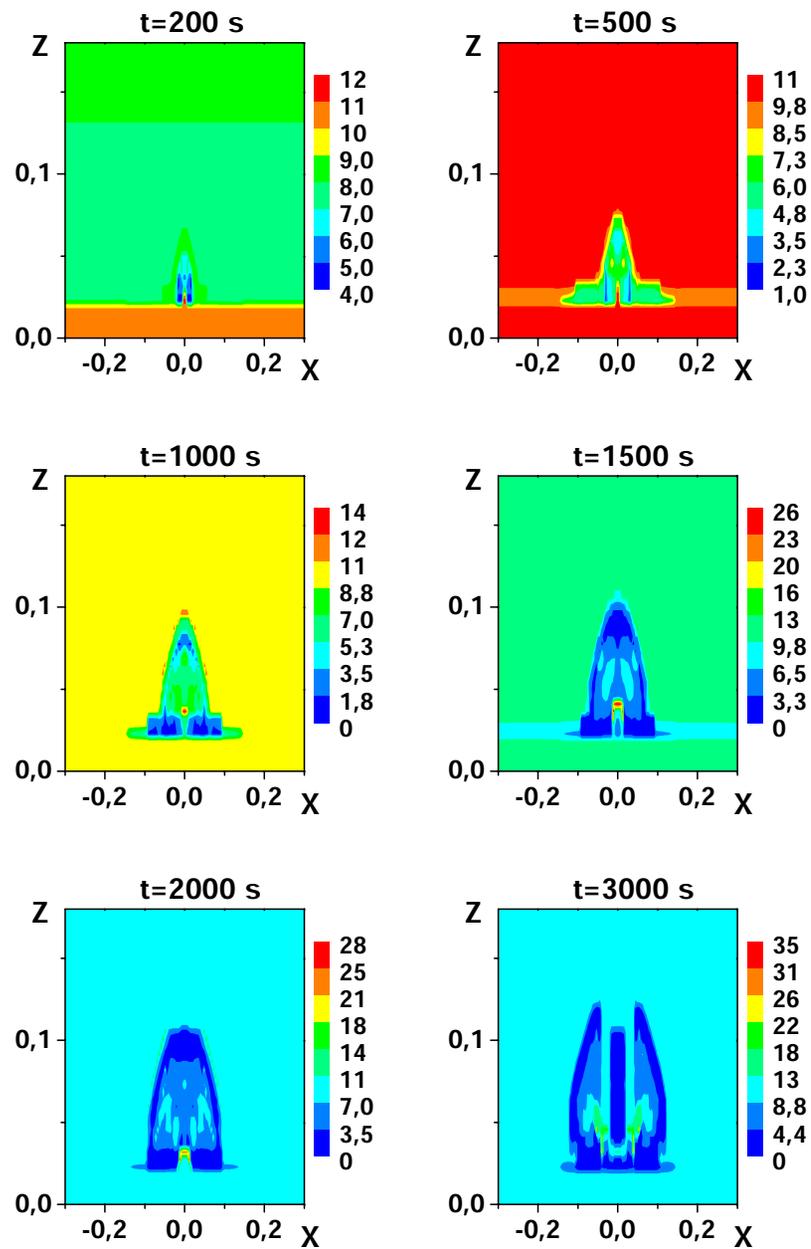
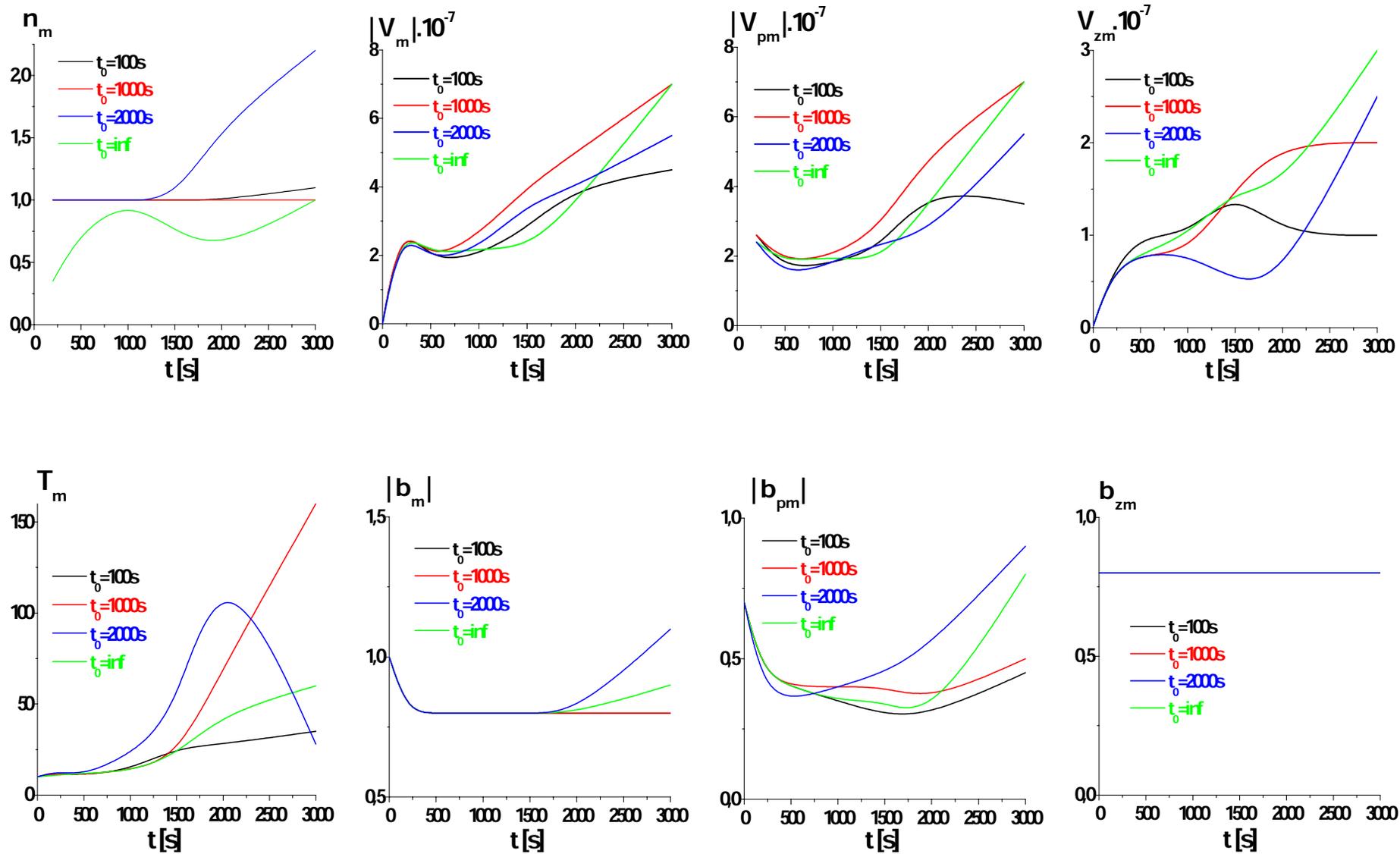
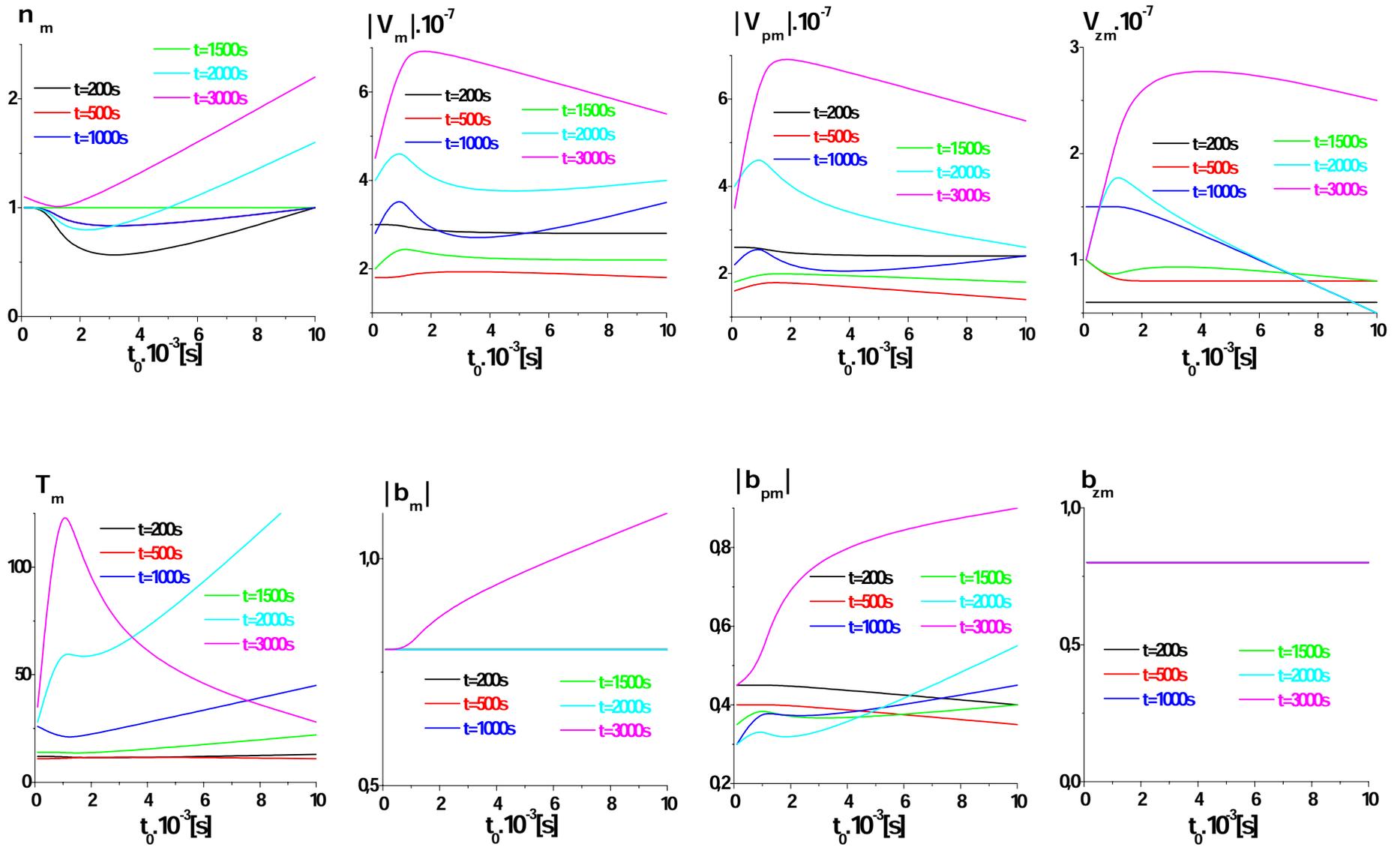
\end{document}